\documentclass[journal]{IEEEtran}

\usepackage{cite}

\ifCLASSINFOpdf
\else
\fi
\usepackage{url}
\usepackage{xcolor}
\definecolor{newcolor}{rgb}{.8,.349,.1}
\usepackage{soul}
\usepackage{amsmath}
\usepackage{algorithm}
\usepackage{algorithmic}
\usepackage{graphicx}
\usepackage{subcaption}
\usepackage{amssymb}
\usepackage{latexsym}
\usepackage{framed,multirow}

\begin{document}
%
\title{Attacking CNN-based anti-spoofing face authentication in the physical domain}
%
%
%

\author{Bowen~Zhang,
		Benedetta~Tondi,~\IEEEmembership{Member,~IEEE,}
        Mauro~Barni,~\IEEEmembership{Fellow,~IEEE}

\thanks{Bowen Zhang is with the School of Cyber Engineering, Xidian University, 266 Xinglong Section of Xifeng Road, Xi’an, Shaanxi 710126, China
 e-mail: (bowenzhang.psnl@outlook.com).}
\thanks{Benedetta Tondi and Mauro Barni are with the Dept. of Information Engineering and Mathematics, University of Siena, Via Roma 56, 53100 - Siena, Italy. 
 e-mail: (benedettatondi@gmail.com, barni@dii.unisi.it)}
}

\maketitle

\begin{abstract}
In this paper, we study the vulnerability of anti-spoofing methods based on deep learning against adversarial perturbations. We first show that attacking a CNN-based anti-spoofing face authentication system turns out to be a difficult task.
When a spoofed face image is attacked in the physical world, in fact, the attack has not only to remove the rebroadcast artefacts present in the image, but it has also to take into account that the attacked image will be recaptured again and then compensate for the distortions that will be re-introduced after the attack by the subsequent rebroadcast process. 
Subsequently, we propose a method to craft robust physical domain adversarial images against anti-spoofing CNN-based face authentication.
The attack built in this way can successfully pass all the steps in the authentication chain (that is,  face detection, face recognition and spoofing detection), by achieving simultaneously the following goals: i) make the spoofing detection fail; ii) let the facial region be detected as a face and iii) recognized as belonging to the victim of the attack.
The effectiveness of the proposed attack is validated experimentally within a realistic setting, by considering the REPLAY-MOBILE database, and by feeding the adversarial images to a real face authentication system capturing the input images through a mobile phone camera.
\end{abstract}


%
\IEEEpeerreviewmaketitle

\section{Introduction}\label{Section:Introduction}
%
%
%
%
\IEEEPARstart{S}{poofing} attacks have become a serious security threat for authentication systems, due to the facility with which they can be used to get unauthorized access to a system by impersonating an authorized user.
In particular, spoofing attacks against face authentication systems can be easily launched by printing a photo of an authorized person on a paper or by displaying it on a digital screen~\cite{evans2019handbookAntiSpoof,li2014OSNfacialdisclosure}.
As a countermeasure, several anti-spoofing techniques have been developed.
Anti-spoofing systems based on deep learning have recently demonstrated their superiority with respect to former methods, and for this reason, they are completely replacing model-based techniques and machine learning methods based on hand-crafted features. For instance, in a competition host by ChaLearn at CVPR, all the 13 teams entering the final round of the competition were adopting a solution based on CNN~\cite{liu2019ChaLearn}.

Despite the very good performance achieved by anti-spoofing systems based on CNNs, likewise in other application domains, CNN methods suffer from the existence of so-called adversarial attacks, a.k.a. adversarial examples, that is, small, often imperceptible, perturbations that, when added to the input of a deep neural network, induce a classification error~\cite{akhtar2018adversarialSurvey}. In a spoofing detection system, this means that an adversarial spoofing face would be misclassified as a real face.
Compared to attacks carried out against CNNs in other application scenarios, however, attacking a CNN-based anti-spoofing system presents a number of significant differences and additional challenges.

\begin{figure}[!h]
\centering\includegraphics[width=3.5in]{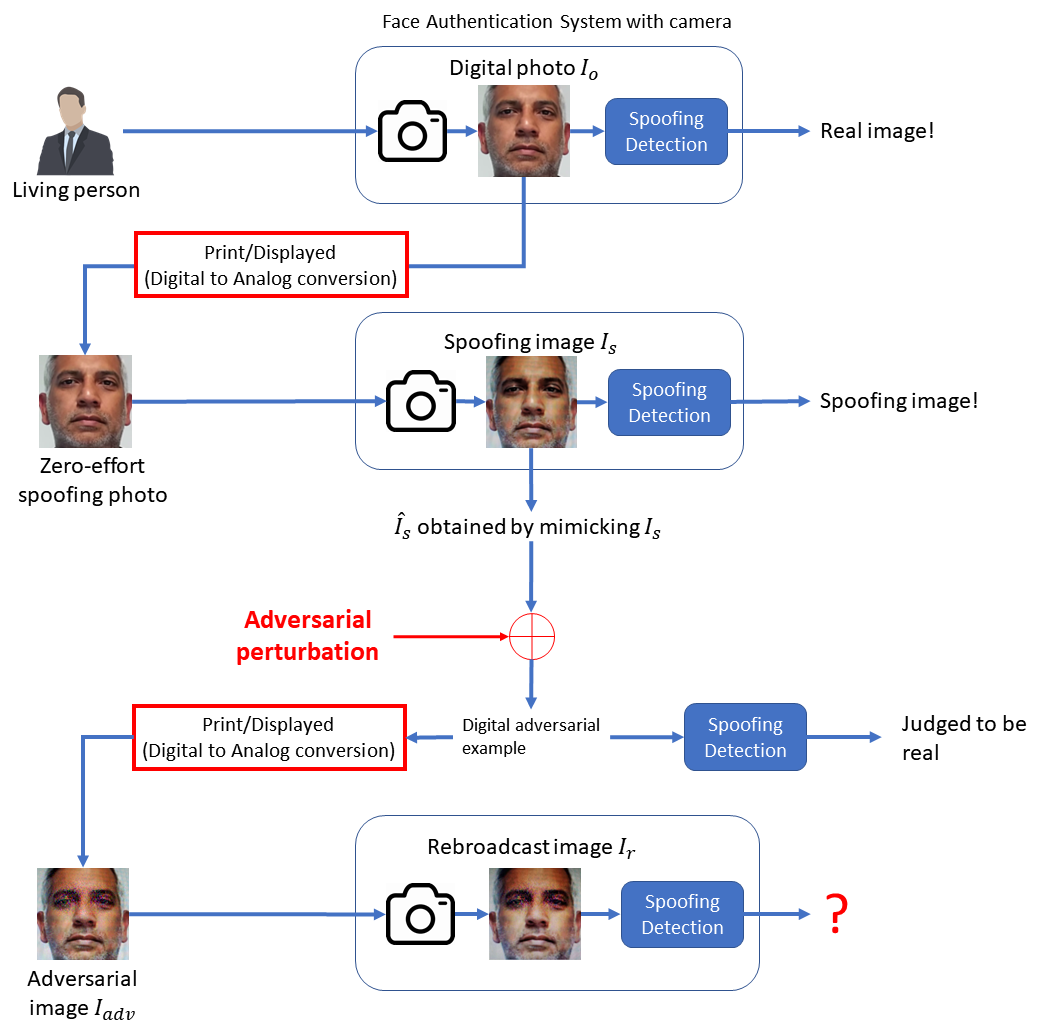}
\caption{Adversarial image generation process against an anti-spoofing system. The process suffers from a double-rebroadcast procedure. To be effective, the adversarial example should fool the system after the second rebroadcast (that is, after the re-capturing of the attacked spoofing image, that can either be printed on a paper or displayed on a screen).} 
\label{fig_1}
\end{figure}

To start with, most works on adversarial attacks assume that the attacker can feed the digitally crafted adversarial example directly into the machine learning model~\cite{szegedy2013intriguing,goodfellow2014explaining,papernot2016JSMA}; such attacks are usually referred to as \emph{digital domain attacks}. However, this assumption does not hold in the case of anti-spoofing, where the system is designed to work in the physical world. 
To clarify this point, let us consider 
a face authentication system equipped with a spoofing detection module, as shown in the first row of Figure~\ref{fig_1}. To access the system, a living person must stand in front of the system, so that a camera takes a digital picture of him/her and sends it to the spoofing detection module, which will classify the image as taken from a real person. The first step of an impersonation attack requires that the attacker retrieves a digital picture of the victim, e.g. the digital photo in the upper row of Figure~\ref{fig_1}~\cite{li2014OSNfacialdisclosure}. The attacker, however, has no access to the internal pipeline of the system. In fact, should the attacker have access to the digital input of the system, the adversarial attack would not be necessary at all, since he could attack the system by simply feeding it with the digital image of the victim. On the contrary, to attack the face anti-spoofing system, the digital image has to be printed/displayed in the physical world and then captured by the system camera, as illustrated in the central part of Figure~\ref{fig_1}. This digital-to-analog, then analog-to-digital process is usually referred to as image or video \textit{rebroadcast}~\cite{agarwal2018rebroadcast}. It is thanks to the image modifications introduced during the rebroadcast process that the anti-spoofing detector can understand that the digital image fed into the system is the result of a spoofing attack and consequently deny the access to the system.
This unavoidable spoofing procedure marks an important difference with respect to other adversarial attacks carried out in the physical domain (usually referred to as \textit{physical domain attacks}~\cite{athalye2017synthesizing,eykholt2018robustroadsign}), because the spoofing detection module looks exactly for the spoofing artefacts introduced during the rebroadcast process, and such artefacts are unavoidably introduced again when a physical-world version of the attacked image is fed into the system, possibly leading to a correct detection of the spoofing attack (see bottom part of Figure \ref{fig_1}). For this reason, the adversarial attacks must act in a pre-emptive way, anticipating that a new spoofing pattern will be introduced again after the atatck.

For the above reasons, creating an adversarial example capable to attack an anti-spoofing system is not easy. The first problem the attacker is faced with, is the choice of the digital image to be used to create the adversarial example. By referring to Figure \ref{fig_1}, the adversarial example can not be created by starting from the original image $I_0$ available to the attacker. Such an image, in fact, would already be judged positively by the anti-spoofing detector, since it has ben obtained by taking a picture of the victim of the attack. In addition, such an image does not contain any traces of the rebroadcast process, so it is of no help for the attacker to build an attack that will work once the attacked image is rebroadcast into the system. A better choice would be to apply the attack to the image $I_s$, obtained by feeding a rebroadcast version of $I_0$ into the system. For this to be possible, however, the attacker should have access to the output of the system camera, and we already argued that is not the case. The solution we consider here, is that the attacker mimic the rebroadcast/recapture process of the anti-spoofing system by building a copy of the target system, say a smart phone, and use it to take a photo in the desired setting. In the following we will refer to such an image as $\hat{I}_s$.
  
Having defined the starting point of the attack, the physical domain adversarial examples must be built in such a way to satisfy the following goals: i) the spoofing  artefacts contained in $\hat{I}_s$ during the rebroadcast process must be removed (to the eyes of the spoofing detection model), ii) $\hat{I}_s$ must be modified in such a way to  preemptively take into account that the attacked image will be rebroadcast and recaptured again, iii) the adversarial perturbation must be robust to the geometric and luminance distortions introduced by the rebroadcast procedure.

Eventually, successfully attacking a face authentication system with anti-spoofing capabilities requires not only that the anti-spoofing module is fooled, but also that the face region is correctly detected by the system and recognized as belonging to the victim of the attack.

In light of the previous discussion, the contribution of this paper can be summarized as follows:
\begin{enumerate}
\item
We show that the straightforward application of off-the-shelf adversarial attacks against a CNN-based anti-spoofing face authentication system does not work.
\item Inspired by the work in \cite{athalye2017synthesizing}, we design a new physical domain attack against a face authentication system. The adversarial spoofing images are crafted in order to achieve the following goals: i) make the spoofing detection fail; ii) the system detects and recognizes the perturbed facial region as a face; iii) the attacked face is recognized by the system as belonging to the victim of the attack.
\item
We experimentally prove the effectiveness of the attack by using it in a realistic, physical world, setting.
\end{enumerate}

The rest of the paper is organized as follows. In Section~\ref{Section:RelatedWork}, we review prior art on CNN-based anti-spoofing and existing attacks against CNNs.  In Section~\ref{Section:FaceAuth}, we describe the end-to-end system targeted in our study, and in Section~\ref{Section:ProposedAttack}, we present the proposed attack. In Section~\ref{Section:experiment setting}, we introduce the methodology we followed in our experiments and the experimental setting. The results of the experiments are provided and discussed in Section~\ref{Section:Experiment}. Finally, we draw our conclusions in Section~\ref{Section:Conclusion}.
\section{Related Work}\label{Section:RelatedWork}
\subsection{CNN-based anti-spoofing}\label{relatedwork-CNNantispoofing}

Motivated by the outstanding performance of deep neural networks (DNNs) on computer vision and pattern recognition tasks, researchers have started exploring the use of DNN methods for spoofing detection.
Yang at al. \cite{yang2014firstCNNonspoofing} first attempted to exploit convolutional neural networks (CNNs) for face spoofing detection; the authors propose to extract self-learned features by using a CNN, and then feed them into an SVM classifier. After that work,  several different methods resorting to CNNs for both feature extraction and classification  have been proposed: Lucena et al. \cite{lucena2017transfer} applied a pre-trained deep VGG net fine-tuned on a spoofing dataset; Li at el. \cite{li2016CNN+PCV}, also used a fine-tuned VGG net and PCA  to reduce the feature dimension before a final SVM classifier. In \cite{atoum2017two-streamCNN}, the authors proposed a two-stream CNN that utilizes both local features and holistic depth maps from face images. 3D CNNs have also been proposed to extract spatial and temporal information simultaneously~\cite{gan20173d,li2018TIFS}.

In general, compared to methods based on hand-crafted features, CNNs can learn more discriminative features, and achieve better performance, at least when they are tested under conditions similar to those used during training.  The use of increasingly rich databases~\cite{liu2019ChaLearn} permits to mitigate the database mismatch problem and improve the generalization capabilities of these solutions. For all these reasons, CNN based methods are leading the trend in anti-spoofing researches.

\subsection{Adversarial Attacks to CNNs}

Szegedy et al. \cite{szegedy2013intriguing} first demonstrated the existence of adversarial examples for deep learning models, i.e., small (often quasi-imperceptible) perturbations that, when added to the input image, yield an incorrect classification.
Such vulnerability is a serious weakness of deep learning models, especially when they are employed in security-related applications. For this reason, the analysis of the robustness of CNN models against adversarial attacks is gaining more and more attention~\cite{akhtar2018adversarialSurvey}.
Following this trend, a number of methods have been proposed to build adversarial examples capable to deceive deep neural networks, including the Fast Gradient Sign Method (FGSM)~\cite{goodfellow2014explaining}, the Basic Iterative Method (BIM)~\cite{kurakin2016adversarial}, the Jacobian-based Saliency Map Attack (JSMA)~\cite{papernot2016JSMA} and many others~\cite{akhtar2018adversarialSurvey}. Specific toolboxes have also been developed, like Foolbox~\cite{rauber2017foolbox} and Cleverhans~\cite{papernot2016cleverhans}, which contain the above as well as other state-of-the-art attacks.

Most of the attacks proposed in the literature, and among them all the attacks mentioned above, share the assumption that the attacker can feed the perturbed digital image directly into the neural network. This kind of attacks are usually referred to as \textit{digital domain attacks}.  In a more realistic scenario, the attackers do not have the possibility to directly feed a digital image into the system; in these cases, the adversarial examples are always captured by cameras or sensors. For example, machine learning systems used for robots vision, video surveillance and face authentication usually get their inputs from video cameras and other sensors. Recently, researchers have investigated the possibility to build adversarial attacks that can work in this more realistic setting. For instance, in~\cite{kurakin2016adversarial} it is shown that adversarial examples can still fool an image classifier after that they are printed and recaptured, assuming that they are presented to the system in an axis-aligned position. At the same time, it has been proven that variations of the viewing distance and angle, quite common in practice, have a non-negligible impact on the performance of attacks carried out in the physical world~\cite{luo2015foveation,lu2017noneed}.

Only a few works have proposed successful attacks in the physical domain, usually referred to as \textit{physical domain attacks}. Sharif et al. \cite{sharif2016accessorize} attacked a face recognition system by printing the adversarial examples on a pair of eyeglass frames. Wearing such glasses would enable an attacker to impersonate a different individual. In particular, the authors look for an adversarial perturbation that can fool the face recognition system for an entire class of images. Such a goal is achieved by optimizing the cross-entropy loss over a set of portrait photos collected by the attacker, which have undergone a set of geometric transformations typical of the recapture process.
This work has shown for the first time that adversarial attacks can be successfully carried out in the real world, even in the presence of geometric distortions and axes misalignment.
In another work ~\cite{athalye2017synthesizing},
the loss function is optimized over several synthetic transformations, that is, by considering the expectation over those transformations. In this way, the authors were able to craft 3D-printed adversarial examples that are robust to perturbations of the viewing distance and angle. Eykholt et al. ~\cite{eykholt2018robustroadsign} proposed a robust physical domain attack to generate adversarial examples that can fool stop sign classifiers. Unlike ~\cite{athalye2017synthesizing}, they applied both synthetic and physical transformations. Later on, such a work has been extended to attack a general object recognition system~\cite{eykholt2018detector}.

All these works share the idea of modelling the possible physical world distortions that the image may be subject to during the optimization procedure, and optimize the average loss over all the distorted images. This approach is also adopted in this paper. Our study, however, deviates significantly from previous works, since part of the distortions introduced during the acquisition process are the traces that the anti-spoofing detection system relies on. So, even if one could create an adversarial perturbation erasing the spoofing patterns present in the image, new patterns would unavoidably be introduced again during the physical domain attack and possibly be detected by the detector. This makes the creation of physical adversarial examples against anti-spoofing networks harder to achieve.

\section{To-be-attacked face authentication system}\label{Section:FaceAuth}

In this section, we describe the end-to-end face authentication system which we are going to attack in our study.

\subsection{Overall structure}
\label{Section:End2endSys}

\begin{figure}[!h]
\centering\includegraphics[width=3.5in]{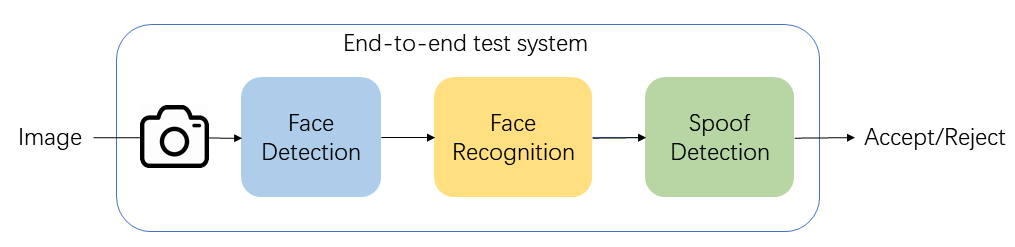}
\caption{Structure of the end-to-end face authentication system considered in this paper.}
\label{fig_2}
\end{figure}

As shown in Figure \ref{fig_2}, the end-to-end face authentication system targeted by our attack consists of a cascade of classifiers whose tasks are: face detection, face recognition and spoofing detection. We observe that the spoofing detection module is the last step of the authentication chain. In this way, the spoofing detection decision is made by analyzing only the facial part of the image. This choice makes attacking the anti-spoofing system harder, since it forces the adversarial perturbation to be restricted to the facial region of the image, with the consequent risk of inducing failure in the face recognition part of the authentication chain.

\subsubsection{Face detection and face recognition networks}\label{Subsection:Face detection and recognition}
In order to implement the face detection and recognition modules, we used the open-source python library \textit{face\_recognition}~\cite{geitgey2017face} . This library is built on top of the open-source library \textit{dlib}~\cite{dlib09}, which provides the CNN model for face detection. The model was trained and tested on the Labeled Faces in the Wild benchmark~\cite{LabelFaceinWild}, and it has a detection accuracy of 99.38\% in the absence of attacks.

For the face recognition module, dlib provides another model that extracts and encodes the features from the faces detected by the face detector network. Following the examples provided by the \textit{face\_recognition} library, we extracted and encoded the features from the enroll-set of REPLAY-MOBILE database~\cite{Replay-Mobile-Costa-Pazo_BIOSIG2016_2016}, the features are further used to train a k-nearest-neighbors (KNN) classifier. The face recognition module was evaluated on the test set and reached 100\% accuracy. We refer to~\cite{dlib09,geitgey2017face} for more information about the CNN structure and other details.

\subsubsection{Spoofing Detection network}\label{Subsec:SpoofingDetection}

\begin{figure}[!h]
\centering\includegraphics[width=3.5in]{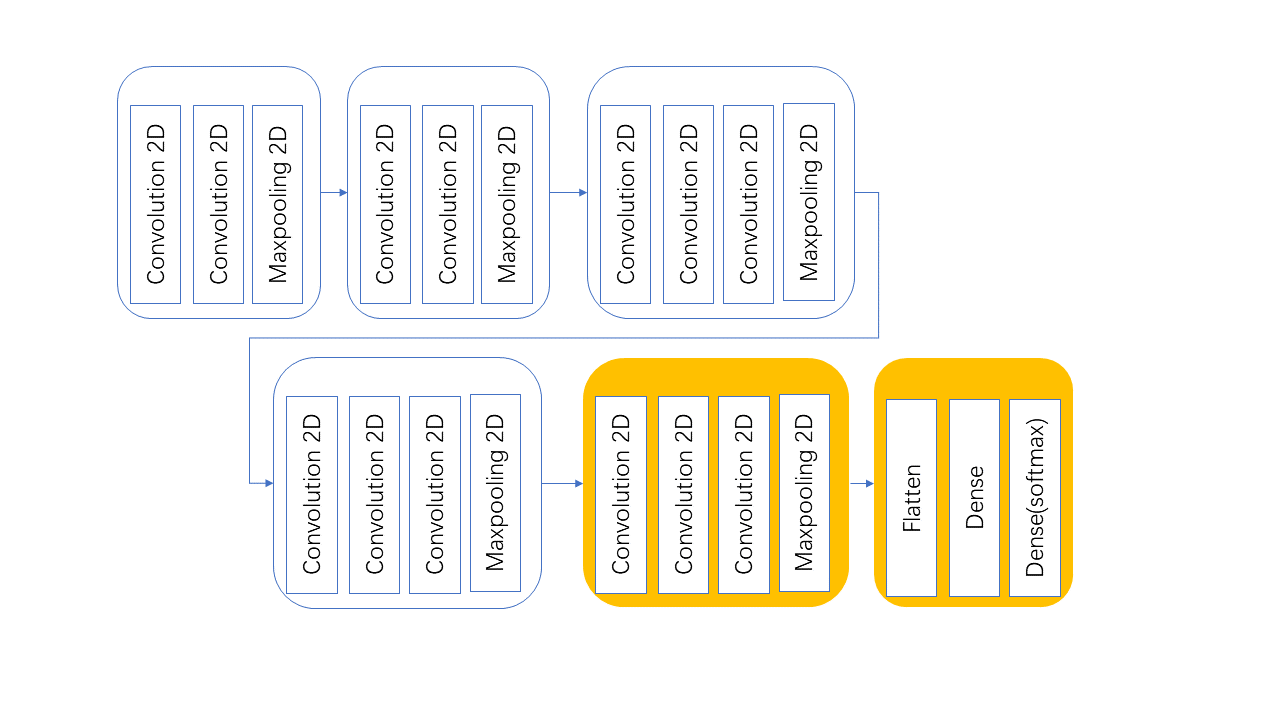}
\caption{Architecture of the spoofing detection CNN. Yellow blocks denote the fine-tuned layers while the weights of the other layers remained fixed.}
\label{finetune}
\end{figure}

\begin{figure}[!h]
\centering\includegraphics[width=3.5in]{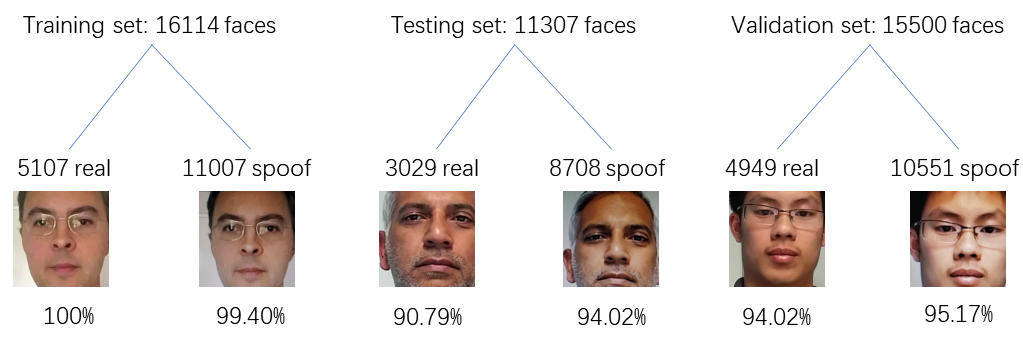}
\caption{Evaluation result of our spoofing detection model on each part of the data set}
\label{datasetPerformance}
\end{figure}

To build the spoofing detection module, we fine-tuned a network pre-trained on the spoofing detection data set~\cite{lucena2017transfer}. Specifically, we took a VGG-16 network pre-trained on ImageNet  database~\cite{russakovsky2015imagenet}. The architecture of the network is shown in Figure \ref{finetune}. The fine-tuned layers are highlighted in yellow in the figure, while the weights of the other layers remained fixed. To fine-tune the network, so to make it work in the setting used for our experiments, we collected our own data set from the same laboratory environment that we used, later on, to carry out the attack experiments. We first took a subset of real-access videos in the REPLAY-MOBILE database~\cite{Replay-Mobile-Costa-Pazo_BIOSIG2016_2016}. Then, we rebroadcasted and recaptured the videos to collect the spoofing class of images. Rebroadcasting was carried out by displaying the original videos on a digital screen, in the mean time a cell phone, fixed on a tripod, was taking pictures of the screen continuously. During the rebroadcasting process, we slowly moved the cell phone to cover different shooting angles and distances. In this way, we obtained about 30,000 spoofing images. At the same time, we took 13,000 frames from the original videos to build the dataset with the original, non-spoofed, images. In the end, the facial part of all the collected images were cropped, and divided into training, testing and validation parts, each containing different identities. The laboratory setting we used to build the
dataset (and then carry out the attack experiments) is depicted
in Figure \ref{fig:labSetting}. For the training process, we used the Adam optimizer with learning rate equal to $10^{-4}$. After 50 epochs, the fine-tuned model achieved 98.39\%, 94.56\% and 93.17\% accuracies on the training, validation and test sets respectively. The details of the performance on each part of the data set are shown in Figure \ref{datasetPerformance}.

\section{Proposed attack}\label{Section:ProposedAttack}

Our goal is to build a physical adversarial attack against the end-to-end system depicted in Figure~\ref{fig_2}, that is, to produce adversarial spoofing images capable to: i) pass the face detection step, ii) be recognized as the individual targeted by the attack, and iii) fool the final spoofing detection check, thus illegally gaining access to the system. Below, we first formulate  the standard problem of generating a digital domain adversarial image, then we refine the formulation to fit the case of physical domain attacks.

Let $I$ denote an input image and $l_{true}$ the corresponding label and let $l_{target}\neq l_{true}$ be the target label of the attack.
Let $f(\cdot)$ denote the soft output of the neural network under attack. In general, finding an adversarial perturbation $\rho$ corresponds to solving the following optimization problem:

\begin{equation}\label{adv likelyhood optimization}
\begin{aligned}
&\operatorname*{arg\,min}_{\mathbf{\rho}}
& & \mathcal{J}(f(I+\mathbf{\rho}),l_{target}),\\
&\text{subject to}
& & \| \mathbf{\rho} \|_p < \epsilon
\end{aligned}
\end{equation}

where $\mathcal{J}(\cdot)$  denotes the loss function of the neural network, and $\|\cdot\|_p$ denotes the $L^p$-norm, the most common choice corresponding to $p=2$. Usually, the above optimization is approximately solved by using the corresponding Lagrangian-relaxed form:
\begin{align}
\label{optimization_problem}
\operatorname*{arg\,min}_{\mathbf{\rho}} \mathcal{J}(f(I+\mathbf{\rho}), l_{target})+\lambda \| \mathbf{\rho} \|_p
\end{align}
where $\lambda$ is a hyper-parameter controlling the strength of the distance penalty term $\| \mathbf{\rho} \|_p$.

In the case of spoofing detection, which is a binary classification problem, the goal of the attacker is to induce the CNN to classify a spoofing image as real. By adopting the formalism introduced in Section \ref{Section:Introduction}  and depicted in Figure \ref{fig_1}, in the sequel we use $\hat{I}_s$ instead of $I$ to denote the spoofing image under attack, therefore the target class $l_{target}$ is always ``real'', with reference to the output of the spoof detection network. For simplicity, in the rest of the paper, we use the label ``0" for ``real" and ``1" for ``spoofing" images.

In a physical domain setting (see Figure \ref{fig_1}), the spoofing detection network is not fed directly with $I_{adv} = \hat{I}_s+\rho^*$ ($\rho^*$ being the optimum perturbation  obtained by solving~\eqref{optimization_problem}), but with its recaptured version $I_r = r(I_{adv}) = r(\hat{I}_s+\rho^*)$, where we use $r{(\cdot)}$ to denote the rebroadcast and recapture procedure. For this reason, there is no guarantee that the perturbation $\rho^*$ will induce a detection error. In fact, we will show in Section \ref{Section:Experiment}, that the spoofing detection network identifies $I_r$ as a spoofed image with high probability.

In general, to obtain a robust perturbation that still works when the image is degraded due to rebroadcast, we need to model the perturbations the image is subject to during the rebroadcast process, and take such a model into account
within the optimization problem in ~\eqref{adv likelyhood optimization} (or ~\eqref{optimization_problem}).
Formally, by denoting with $\cal{R}$  the set of distortions the attack must be robust to, the perturbation $\rho$ is found by optimizing the average loss over $\cal{R}$, that is,
\begin{align}\label{physical optimization}
\operatorname*{arg\,min}_{\mathbf{\rho}} \mathbb{E}_{r\sim \cal{R}} [ \mathcal{J}(f(r(I+\mathbf{\rho})),l_t) ]+\lambda \| \mathbf{\rho} \|_p.
\end{align}
The above formulation has been first introduced in~\cite{athalye2017synthesizing}, in the attempt to get an adversarial perturbation robust to geometric image transformations, such as
angle and viewpoint perturbations and is referred to as Expectation over Transformation (EOT). The effectiveness of such an approach has been validated in~\cite{athalye2017synthesizing} in an object recognition context.
By following the EOT approach, a robust adversarial physical domain attack capable to fool the end-to-end face authentication system in Figure \ref{fig_2} can be obtained by solving the following optimization:

\begin{align}
\label{physical optimization2}
\operatorname*{\min}_{\mathbf{\rho}}  \hspace{0.25cm} & \mathbb{E}_{r\sim \cal{R}} [ \mathcal{J}(f_s( r(\hat{I}_s+\mathbf{\rho})),l_t)] +\lambda \| \mathbf{\rho} \|_p\nonumber\\
& s.t. \hspace{0.1cm} \phi(f_d( r(\hat{I}_s+\mathbf{\rho}))) = 1, \phi(f_r( r(\hat{I}_s+\mathbf{\rho}))) = p_{\hat{I}_s},
\end{align}
where $\phi()$ is the $argmax$ function used to obtain the predicted label from the soft output, $f_s$ denotes the soft output of the spoofing detection network, $\mathcal{J}$ is its loss function and $f_d$, $f_r$ denote the decision functions of the face detection
and face recognition networks respectively. For the face detector $\phi(f_d(\cdot))=1$ stands for ``face" (and $0$ for ``non face"). For the face recognition network $\phi(f_r(\cdot))=p_{\hat{I}_s}$ denotes the correct identity for the image $\hat{I}_s$.
The attack is successful when $\phi(f_s(r(\hat{I}_s+\rho^*)))=0$ subject to the two constraints.

We now investigate what happens when the adversarial example is used against a spoofing detection system. A well-trained spoofing detection network should be able to distinguish an original image $I_o$ and its rebroadcast version $\hat{I}_s = r(I_o)$, for any $r\in \cal{R}$ (the set $\cal{R}$ is usually defined by the way the training set is built). For a valid physical adversarial perturbation, we should have $\phi(f_s(r'(r(I_o)+\rho)))=0$ for $r',r\in \cal{R}$. This is a hard goal to achieve, since the network $f_s()$ has just been trained to provide a large response on any $r''(\cdot) \in \cal{R}$.
Given the above observation, we need to characterize $\cal{R}$ in the anti-spoofing setting targeted in this paper, in order to build an attack which is at the same time robust to geometric perturbations and robust to the degradation caused by the rebroadcast procedure.

\subsection{Rebroadcast modeling}

As indicated in previous works~\cite{athalye2017synthesizing,eykholt2018robustroadsign,jourabloo2018DeSpoofing}, the degradation introduced by rebroadcasting is due to two main factors: geometric distortions, including rotation, view distance and angle changes, and displaying or printing artefacts, like color distortion and light reflection.
In our work, we decided to model the geometric distortion by simple transformations like rescaling, translation, and perspective transforms, and the display artefacts by means of hue, saturation, contrast and brightness changes. A complete description of all the transformations that we used to model the rebroadcast artefacts is given in Table~\ref{transformations}, where the range of the parameters characterizing each transformation is also reported. The transformations were implemented by using the open-source python library imgaug, version 0.2.8~\cite{jung4imgaug}, to which we refer for a more detailed explanation of the transformation parameters.

For EOT, every $r()$ in \eqref{physical optimization} (and \eqref{physical optimization2}) is a composite transformation consisting of all the transformations in Table \ref{transformations} applied in random order. Every single transformation is applied with the parameters randomly selected in the allowed range. The average loss is then computed over 2000 versions of the to-be-attacked image obtained by applying to it the composite transformations.

\begin{table}[h]
\centering
\caption{Distortions used to model the rebroadcast process.}
\label{transformations}
\begin{tabular}{|c|c|c|}
\hline
\multicolumn{2}{|c|}{Trasformation}      & Range                   \\ \hline
\multirow{4}{*}{Affine}   & Rotation     & $[-5^\circ, 5^\circ]$   \\
                          & Shear        & $[-5^\circ, 5^\circ]$   \\
                          & Scaling      & $[0.85,1.15]$           \\
                          & Translation  & $[0,15\%]$of image size \\ \hline
\multicolumn{2}{|l|}{Perspective}        & $[0,0.025]$             \\ \hline
\multicolumn{2}{|l|}{Brightness}         & $[0.85,1.15]$           \\ \hline
\multicolumn{2}{|l|}{Constrast}          & $[0.9,1.1]$             \\ \hline
\multicolumn{2}{|l|}{Gaussian Blurring(stdev)}  & \multicolumn{1}{c|}{$[0,1]$}       \\ \hline
\multicolumn{2}{|p{0.5\columnwidth}|}{Hue and Saturation (value added to H and S Channel)} & \multicolumn{1}{c|}{$[-15,15]$}       \\ \hline
\end{tabular}
\end{table}

\subsection{Solving the attack optimization problem}

Previous works have shown that the optimization problem in~\eqref{physical optimization} can be solved by well-know gradient descent algorithms like Adam~\cite{kingma2014adam} or projected gradient descent. In our case, an additional difficulty could be represented by the presence of the two additional constraints in~\eqref{physical optimization2}, however we found that solving \eqref{physical optimization} with an additional condition on the Peak Signal-to-Noise Ratio (PSNR) is enough. The reason for such an observation is the following.
Even if adversarial examples maintain part of their effectiveness against models other than that targeted by the attack, this property (often referred to as attack transferability) does not generally hold when adversarial examples are transferred between models with different tasks~\cite{papernot2016transferability}. Therefore our attack against the spoofing detection model is likely to have a limited effect on the face detection and recognition networks. Yet, we still need to limit the distortion introduced by the attack to avoid that the attacked image is so degraded to impede a correct operation of the face detection and image recognition modules. For this reason, we set a PSNR stop condition to avoid such an extreme case. In fact, we verified experimentally that lower-bounding the PSNR to 20dB, the attacked images pass the face detection and recognition steps (see Section~\ref{Section:Experiment}).

\begin{figure}[ht]
	\centering
	~ 
	\begin{subfigure}[b]{\linewidth}
		\centering
		\includegraphics[width=2.5in]{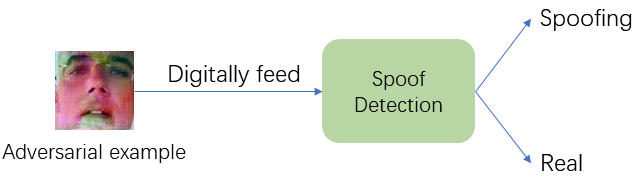}
		\caption{Digital domain test pipline}
		\label{fig:digitalTest}
	\end{subfigure}
	~ 
	\begin{subfigure}[b]{\linewidth}
		\includegraphics[width=3.5in]{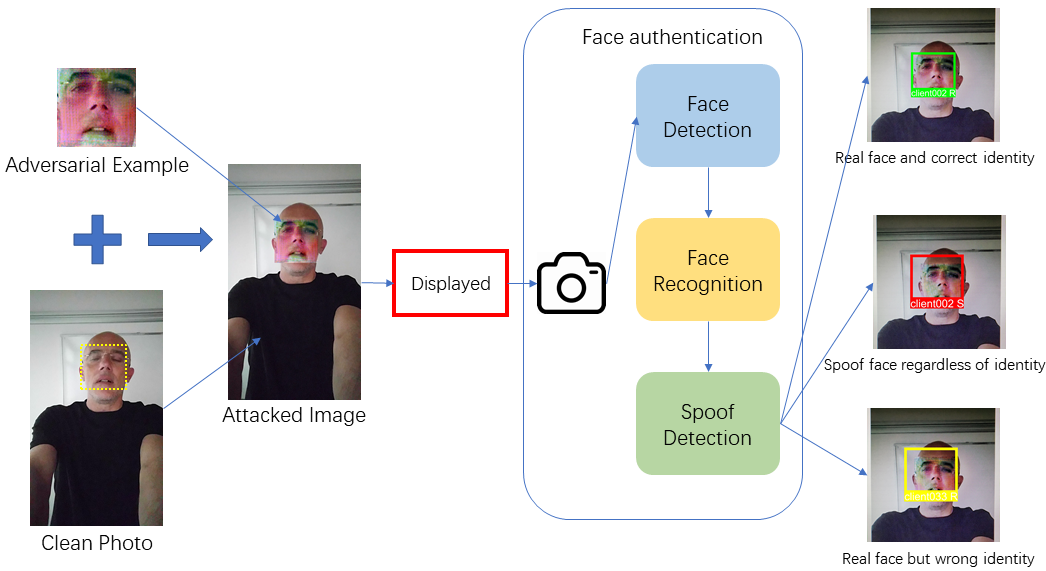}
		\caption{Physical domain test pipline}
		\label{fig:physicalTest}
	\end{subfigure}
	\caption{Experimental pipelines to evaluate the performance of attacks.}
\label{testPipeline}
\end{figure}

As we said, to test the effectiveness of our attack in a general and practical way, we considered a face authentication system that takes as input pictures containing both face and background, while the spoofing detection module considers only the small region containing the facial part only (as detected by the face detection module). Yet, in the previous sections, the adversarial attacks (be them performed in the physical or digital domain) are applied only to the facial region of the image. Then, to test the adversarially perturbed face against the entire authentication system, the attacked facial region is embedded into an overall picture containing the face, part of the body, and the background. The resulting image is simply denoted as \textit{attacked image}, and is used to launch the spoofing attack against the end-to-end system. Figure \ref{testPipeline} shows an example of adversarial examples against the spoofing detection network and the correspond attacked image.

\section{Experimental methodology}\label{Section:experiment setting}

\subsection{Experimental setting}
\label{expset}

\begin{figure}[!h]
	\centering\includegraphics[width=0.8\columnwidth]{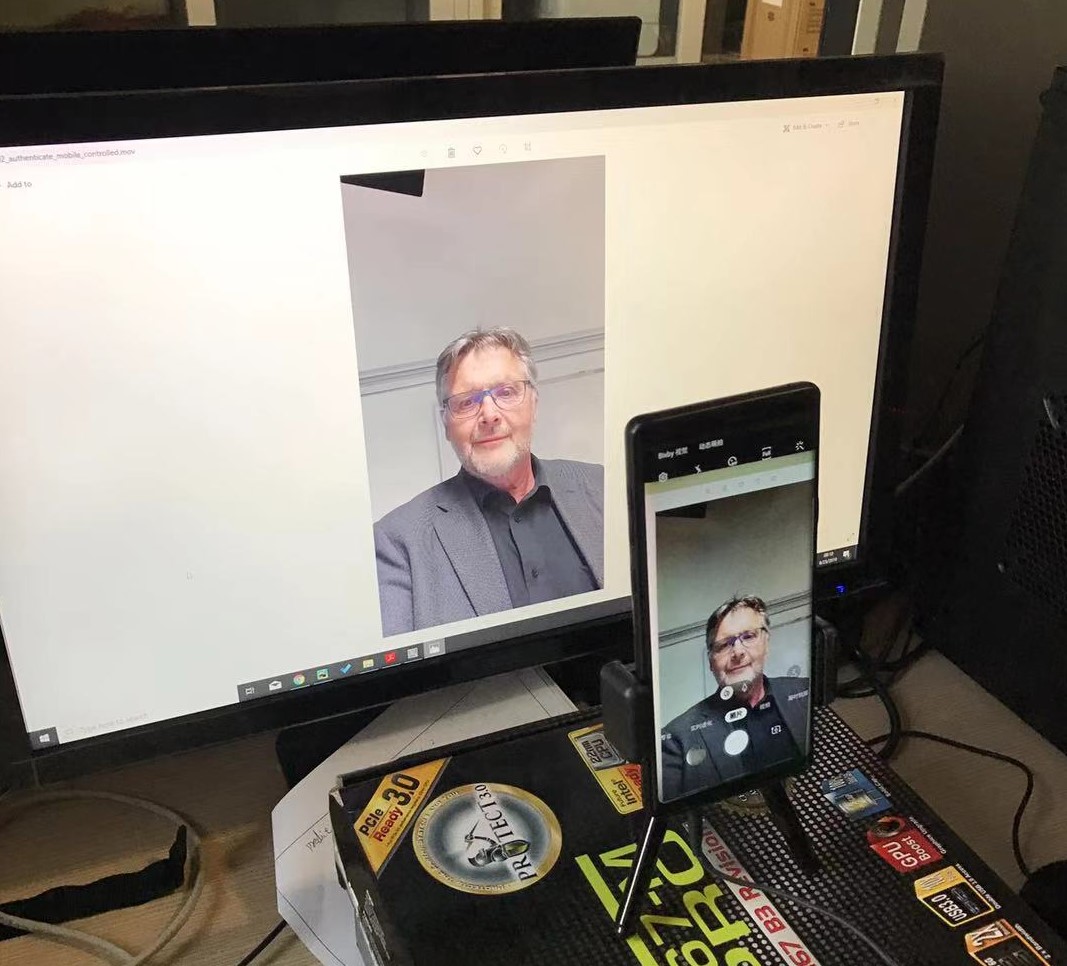}
	\caption{Laboratory setting used to build fine-tuning data set and perform attack experiments.}
	\label{fig:labSetting}
\end{figure}

For our experiments, we used the same laboratory setting used for the dataset collection described in Section~\ref{Subsec:SpoofingDetection} (see Figure \ref{fig:labSetting}).
We used an Acer KA240HQ digital screen (23.6 inches, 1920 $\times$1080 resolution) as the monitor to display the spoofed images, and a Sumsung Note9 cell phone as the acquisition camera (12.0 MP resolution). The shooting angle and distance in all the experiments and during dataset collection (Section \ref{Section:FaceAuth}) range from -20$^\circ$ to 20$^\circ$, and from 10cm to 50 cm, respectively.

To show that the direct rebroadcast of an image attacked in the digital domain is not enough to fool the spoofing detection system, we used several digital attacks included in the  \textit{foolbox}~\cite{rauber2017foolbox} library (version  1.8.0), that is, the Fast Gradient Sign Method (FGMS) introduced in ~\cite{goodfellow2014explaining}, the Basic Iterative Method (BIM) in~\cite{kurakin2016adversarial}, the Gradient Attack (GA), Iterative Gradient Attack (IGA), and the Iterative Gradient Sign Attack (IGSA). The GA  is a one-shot implementation of the standard gradient attack. The IGA and IGSA are the iterative versions of GA and FGSA, respectively,  which, even if more time consuming, are more effective than the single-step attack methods. For each attack, we generated two sets of adversarial examples using different parameters, introducing perturbation with different strengths. For reproducibility, we list the parameters of the attacks and the average PSNR of the adversarial examples in Table~\ref{Table:FoolboxPara}\footnote{For each attack, foolbox provides many parameters that help the user to control the strength of the attack. Giving and explanation of all such parameters would require too much space thus, for simplicity, we only give the values of the parameters and refer to \cite{rauber2017foolbox} for their meaning}. 

With regard to the proposed physical domain attacks, we also considered two sets of adversarial examples. The first set was obtained by setting  $\lambda = 0$ in \eqref{physical optimization2}, that is without considering the distance penalty in the loss function (Set\#1, average PSNR=21.97). In this case, the effectiveness of the attack is maximized, at the price of a stronger distortion. To build the second set, we set the penalty term to $\lambda = 0.1$  (Set\#2, average PSNR=24.73) so to penalize larger distortions.

To mimic the acquisition part of the attacked system and forge the to-be-attacked images $\hat{I}_s$, we used another Sumsung Note9 with the same characteristics of the one used in the face authentication system. In total, we forged 160 spoofing images, corresponding to 9 different identities. We verified that all of the them are correctly classified as spoofing images when they are rebroadcast and used as input of the spoofing detection network. These 160 images were used to generate all the adversarial examples used in our experiments, including 5 digital domain attacks and the proposed physical domain attack.

\begin{table}[h]
	\caption{Parameters used for existing digital domain attacks. Each row corresponds to one set of adversarial examples. ([0.1:1.1,1000] denotes 1000 values evenly distributed from 0.1 to 1.1, ``\textbackslash{}" denotes the parameter is not available for this attack.}
	\label{Table:FoolboxPara}
	\begin{tabular}{|c|c|c|c|c|}
		\hline
		& epsilon              & step\_size       & \begin{tabular}[c]{@{}c@{}}binary\\  \_search \end{tabular}   & PSNR  \\ \hline
		\multirow{2}{*}{BIM}  & 0.3                  & default          & Flase            & 21.25 \\ \cline{2-5} 
		& 0.7                  & 0.08             & False            & 25.46 \\ \hline
		\multirow{2}{*}{FGSM} & {[}0.1:1.1,1000{]}   & \textbackslash{} & \textbackslash{} & 20.87 \\ \cline{2-5} 
		& {[}0.05:0.5,1000{]}  & \textbackslash{} & \textbackslash{} & 26.01 \\ \hline
		\multirow{2}{*}{GA}   & {[}0.1:1.1,1000{]}   & \textbackslash{} & \textbackslash{} & 21.93 \\ \cline{2-5} 
		& {[}0.05:0.5,1000{]}  & \textbackslash{} & \textbackslash{} & 26.73 \\ \hline
		\multirow{2}{*}{IGSA} & {[}0.075:0.75,100{]} & \textbackslash{} & \textbackslash{} & 21.90 \\ \cline{2-5} 
		& {[}0.05:0.5,100{]}   & \textbackslash{} & \textbackslash{} & 25.57 \\ \hline
		\multirow{2}{*}{IGA}  & {[}0.08,0.8,100{]}   & \textbackslash{} & \textbackslash{} & 21.17 \\ \cline{2-5} 
		& {[}0.05:0.5,100{]}   & \textbackslash{} & \textbackslash{} & 25.47 \\ \hline
	\end{tabular}
\end{table}

\subsection{Evaluation metrics}\label{Subsec:EvaluationMetric}

The effectiveness of the various attacks is measured by means of the \textit{Attack Success Rate} (ASR), namely,  the ratio of the number of attacked images that fool the system to the total number of images being attacked. We evaluated all the attacks (including the digital domain attacks and the proposed attack) in both the digital domain and the physical domain, each time following the corresponding pipeline. 

The performance in the digital domain are assessed with respect to the spoofing detection network targeted by the attack. As shown in Figure \ref{fig:digitalTest}, the digital adversarial examples are directly fed into the spoofing detection model, then, the attack is successful when the adversarial spoofing face is detected as real, i.e., $\phi(f_s(I_{adv})) = 0$.

In the physical domain,  digital adversarial examples are first embedded into the whole picture to form the attacked images. The attacked image is displayed on the monitor, captured by the camera and finally passed through the three modules of the authentication system. The performance are assessed with respect to the entire face authentication system, that is, by measuring the ratio between the number of images detected as real, non-spoofed, faces and recognized as belonging to the victim of the attack, and the total number of tests. An example of successful attack is reported in Figure \ref{fig:physicalTest}. For sake of clarity, we use the symbol ASR$_D$ to denote the success rate when the attack is carried out in the digital domain and ASR$_P$ when the attack is carried out in the physical domain.


\section{Experimental Results}\label{Section:Experiment}

We first show that digital domain attacks designed without taking into account the recapture process fail when they are applied {\em as they are} in a physical world scenario. Then, we demonstrate the effectiveness of the proposed physical domain attack.

\subsection{(In)effectiveness of digital domain attacks  in a physical domain setting}

We tested the performance of the 5 existing digital attacks described in Section \ref{expset} when they are applied in the physical domain. The results we got are in Table \ref{PSNR20} (PSNR $\simeq$ 20dB) and Table \ref{PSNR25} (PSNR $\simeq$ 25dB).  All the attacks have a good performance when they are applied in the digital domain. In particular, the iterative attacks can always reach 100\% ASR$_D$ under all PSNR conditions. GA and FGSM are suboptimum attacks, then, not surprisingly, the ASR$_D$ is lower. In all cases, the performance drop significantly when the attacks are applied in the physical domain, and nearly all of them fail with an ASR$_P$ lower than 30\%. Expectedly, the stronger attacks (PSNR $\simeq$ 20dB) exhibit a higher success rate, which, however, remains always below 50\%.

\begin{table}[h]
\caption{Performance of digital attacks with PSNR limitation set to 20dB.}
\label{PSNR20}
\begin{tabular}{|c|c|c|c|}
\hline
             & PNSR  & \begin{tabular}[c]{@{}c@{}}ASR$_D$\\  in digital domain\end{tabular} & \begin{tabular}[c]{@{}c@{}}ASR$_P$\\ in physical domain\end{tabular} \\ \hline
BIM          & 21.25 & 100\%                                                                            & 42.24\%                                                                      \\ \hline
FGSM         & 20.87 & 85\%                                                                             & 26.99\%                                                                      \\ \hline
GA           & 21.93 & 77.20\%                                                                          & 35.99\%                                                                      \\ \hline
IGSA         & 21.90 & 100\%                                                                            & 32.5\%                                                                       \\ \hline
IGA          & 21.17 & 100\%                                                                            & 39.53\%                                                                      \\ \hline
\end{tabular}
\end{table}

\begin{table}[h]
\caption{Performance of digital attacks with PSNR limitation set to 25dB.}
\label{PSNR25}
\begin{tabular}{|c|c|c|c|}
\hline
             & PNSR  & \begin{tabular}[c]{@{}c@{}}ASR$_D$\\  in digital domain\end{tabular} & \begin{tabular}[c]{@{}c@{}}ASR$_P$\\ in physical domain\end{tabular} \\ \hline
BIM          & 25.46 & 100\%                                                                            & 28.48\%                                                                      \\ \hline
FGSM         & 26.01 & 87.29\%                                                                          & 8.77\%                                                                       \\ \hline
GA           & 26.73 & 86.32\%                                                                          & 18.88\%                                                                      \\ \hline
IGSA         & 25.57 & 100\%                                                                            & 13.75\%                                                                      \\ \hline
IGA          & 25.47 & 100\%                                                                            & 27.38\%                                                                      \\ \hline
\end{tabular}
\end{table}

\subsection{Effectiveness of the proposed attack}
In this section, we assess the performance of the proposed physical domain attack.
For the sake of completeness,  we first show how the attack performs in the digital domain, then we move to the physical domain scenario.
We also assess the performance of the proposed attack in a realistic testing scenario, where we assume that the authentication system can be queried multiple times from the same user before the access to the system is blocked (as it is the case with most practical systems). In this scenario, the attack effectiveness is expected to improve since the attack can be performed multiple times in the attempt to impersonate the targeted user. This behaviour is confirmed by the results we got.

\subsubsection{Effectiveness in the digital domain}

By using the setting described in Section \ref{Section:experiment setting}, we verified that all the 320 adversarial examples of Set\#1 and Set\#2 fool the spoofing detection network when they are fed to the system directly in the digital domain, with 100\% ASR$_D$ for both Set\#1 and Set\#2. Such results are provided in Table \ref{PhyAdvinDigitalDomain} , together with the average PSNR for the two sets.

\begin{table}[h]
\caption{Performance of the proposed attack.}
\label{PhyAdvinDigitalDomain}
\begin{tabular}{|c|c|c|c|}
\hline
\begin{tabular}[c]{@{}c@{}}Adversarial\\examples\end{tabular} & \begin{tabular}[c]{@{}c@{}}Average\\ PNSR\end{tabular} & \begin{tabular}[c]{@{}c@{}}ASR$_D$ in\\ digital domain\end{tabular} & \begin{tabular}[c]{@{}c@{}}ASR$_P$\\  in physical domain\end{tabular} \\ \hline
Set\#1                                                                        & 21.97                                                  & 100\%                                                           & 82.5\%                                                                     \\ \hline
Set\#2                                                                         & 24.73                                                  & 100\%                                                           & 75\%                                                                     \\ \hline

\end{tabular}
\end{table}

\begin{figure}[]
\centering
\includegraphics[width=3.5in]{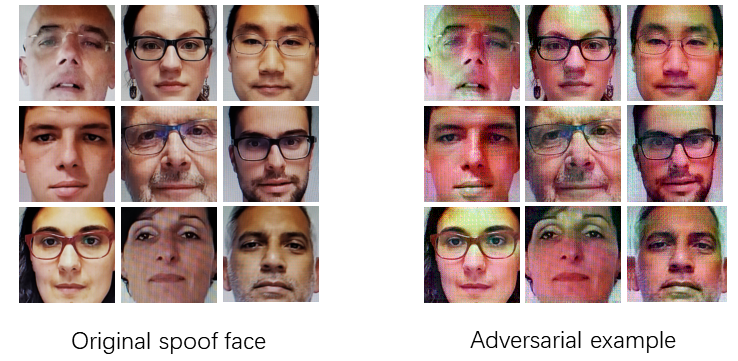}
\caption{Selected adversarial examples. }
\label{advexample}
\end{figure}

\subsubsection{Effectiveness in the physical domain}

\begin{figure}
    \centering
    ~ 
    \begin{subfigure}[b]{.5\textwidth}
        \includegraphics[width=3.5in]{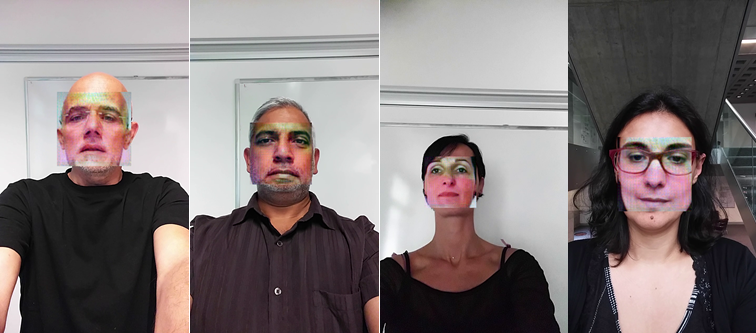}
        \caption{With misalignment}
        \label{fig:misalignment}
    \end{subfigure}
    ~ 
    \begin{subfigure}[b]{.5\textwidth}
        \includegraphics[width=3.5in]{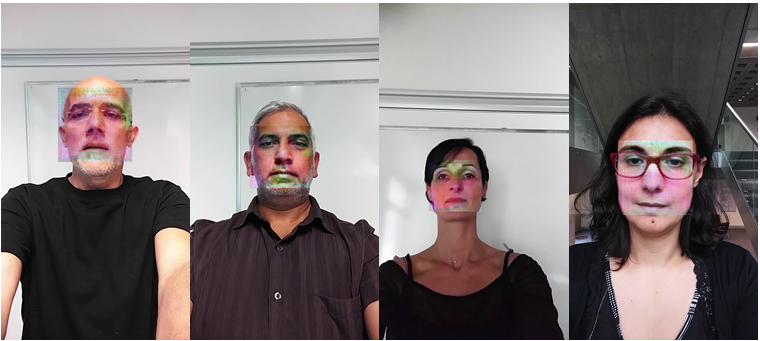}
        \caption{Without misalignment}
        \label{fig:nomisalignment}
    \end{subfigure}
    \caption{Attacked images from Set\#1: (a) attacked images without re-alignment and (b) with realignment. }\label{fig:attackedimages}
\end{figure}

When we pass to the physical domain, as expected, the performance drop a bit (last column in Table \ref{PhyAdvinDigitalDomain}). However, unlike with digital attacks, the ASR$_P$ remains good. Not surprisingly, the best performance in the physical setting are obtained by the images in  Set\#1 with an ASR$_P$ = 82.5\%.
The performance remain reasonably good (ASR$_P$ = 75\%)  even if we decrease the strength of the perturbation as done in Set\#2.
Some examples of attacked images from Set\#1 are shown in Figure \ref{advexample} and Figure \ref{fig:misalignment}.

\begin{figure}[]
\centering\includegraphics[width=3.5in]{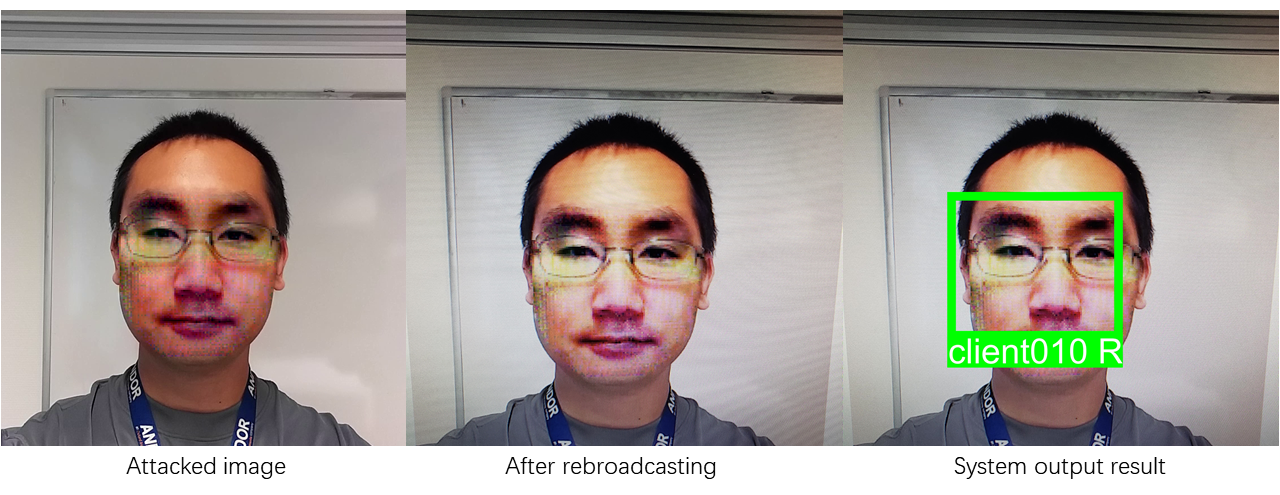}
\caption{Attacked image post-processed with Photoshop.}
\label{PSlevel}
\end{figure}

With regard to the visual quality of the attacked images, we notice from Fig.  \ref{fig:misalignment} that some adversarial images show a slight misalignment between the face and the background. This is because no extra alignment is applied when the attacked face is re-introduced into the overall image.
To mitigate this phenomenon, a possibility is to post-process the images. We did that and manually crafted 100 attacked images from Set\#1.
As shown in Figure \ref{fig:nomisalignment}, these new adversarial images look more natural. The end-to-end tests showed that this mismatch does not affect much the final result, and 80\% of the manually aligned adversarial images successfully fool the system.
This is an important result, since one may argue that a refined spoofing detector could take advantage of the misalignment to detect the presence of the attack. As to the feasibility of manual realignment, it is reasonable to argue that the  attacker can spend some time to craft better pictures, e.g. carefully blurring and feathering the images, adjusting the edges through blending, etc \dots.
%
Figure \ref{PSlevel} provides  an example of a visually pleasant attacked image achieved by adjusting its quality with Photoshop. The image can still fool the spoofing detection system.


It is worth to observe that the majority of rejections given by the authentication system comes from the spoofing detection module. Specifically,  all the attacked images in Set\#1 and Set\#2
pass the face detection module. Regarding the face recognition step, only a few attacked images occasionally failed at this level, in which case the detected face is mistakenly recognized as belonging to an individual other than the target one. Table \ref{failureDistribute} details the distribution of attack failures across the various steps of the authentication system.

\begin{table}[]
\caption{Distribution of attack failures across the various steps of the authentication system}
\label{failureDistribute}
\begin{tabular}{|l|l|l|l|}
\hline
        & face detection & face recognition & spoofing detection \\ \hline
Set\#1  & 0\%            & 7.14\%           & 92.86\%            \\ \hline
Set\#2  & 0\%            & 5\%              & 95\%               \\ \hline
\end{tabular}
\end{table}

\begin{table*}[bt]
\centering
\caption{ISR for the different identities in the two sets of attacked images. For each identity, the upper row denotes the result of Set\#1, and the lower row that of Set\#2. Each entry is given shown in the ``$(N_f,\,N_t)=ISR$" form.}

\label{ISR table}
\begin{tabular}{|c|c|c|c|c|c|}
\hline
Identity                   & ISR             & Identity                   & ISR            & Identity                   & ISR            \\ \hline
\multirow{2}{*}{Client01} & (0,20)=100\%    & \multirow{2}{*}{Client02} & (9,20)=92.63\% & \multirow{2}{*}{Client03} & (2,20)=100\%   \\ 
                           & (0,20)=100\%    &                            & (6,20)=98.25\% &                            & (4,20)=99.65\% \\ \hline
\multirow{2}{*}{Client04} & (2,20)=100\%    & \multirow{2}{*}{Client05} & (2,20)=100\%   & \multirow{2}{*}{Client06} & (0,10)=100\%   \\ 
                           & (9,20)=92.63\%  &                            & (0,20)=100\%   &                            & (0,10)=100\%   \\ \hline
\multirow{2}{*}{Client07} & (0,20)=100\%    & \multirow{2}{*}{Client08} & (12,20)=80.7\% & \multirow{2}{*}{Client09} & (1,10)=100\%   \\ 
                           & (10,20)=89.47\% &                            & (5,20)=99.12\% &                            & (6,10)=83.33\% \\ \hline
\end{tabular}
\end{table*}

\subsubsection{Test scenario with multiple attacks}
Practical authentication systems allow several attempts from the same user before blocking the access. In this setting, an attacker is successful if he can pass the authentication system within the allowed number of access attempts.
To elaborate, let us assume that the authentication system allows at most 3 access attempts, and define a new measure, named \textit{Impersonate Success Rate} (ISR), which is the probability that the attacker succeeds to impersonate a given user at least once. The average ISR over all the identities gives the attack success rate in this scenario with multiple trials.

For a given identity, suppose the attacker has $N_t$ adversarial examples (with the corresponding attacked images) obtained starting from different real images. Let $N_f$  be the number of attack failures out of $N_t$($N_f \le N_t$), i.e. ASR$_P$ =1-$N_f / N_t$.  The ISR for one identity  can then expressed  as:\footnote{We assume that the attack will chose randomly  3 out of the $N_t$ attacked spoofing images for the choosen identity (since he does not know which are the most effective attacked samples).}
\begin{align}
ISR=& 1-\frac{N_f}{N_t}\left(\frac{N_f-1}{N_t-1}\right)\left(\frac{N_f-2}{N_t-2} \right) =1-\frac{\binom{N_f}{3}}{{\binom{N_t}{3}}},
\end{align}
where $\binom{n}{k}$ denotes the binomial coefficient.

The ISR of each identity in the two sets of adversarial examples is shown in Table \ref{ISR table}. For each identity, the upper row denotes the result of Set\#1, and the lower row that of Set\#2.  As one can see, a large ISR is achieved for all identities, and the ISR reaches 100\% in 7 out of 9 cases, with the worst case being ISR = 80.7\%.
The average ISR over all the identities is 97.04\% and 95.82\% for Set \#1 and \#2, respectively.

\section{Conclusions}\label{Section:Conclusion}

In this paper, we have built a physical domain attack against a CNN-based face authentication systems equipped with an anti-spoofing module.
We first showed that, compared to other application scenarios where the attacks are carried out in the physical domain, performing an attack against an anti-spoofing face authentication systems is more difficult and poses a number of additional challenges.
By carefully crafting the attack, and by paying particular attention to modelling the distortions introduced by the rebroadcast process, we have shown that such an attack is indeed possible. In particular, the experimental results we have got confirm that, by using our method, it is possible to successfully craft adversarial spoof images that can at the same time pass the spoofing detection check, the face detection check, and deceive the recognition system (the attacked face is recognized as belonging to the victim of the attack).

Future work may focus on extending our attack to a scenario in which the attack is carried by showing a printed version of the adversarial example to the authentication system. The case of face authentication based on video data is also worth investigation. From the defender's side, future research has to be performed to recover robustness against anti-spoofing, and design new CNN-based face authentication systems capable to work in the presence of adversarial spoofing attacks.

\section*{Acknowledgments}
This work has been partially supported by the China Scholarship Council(CSC), file No.201806960079. The author would also thank Prof. Xixiang Lv (xxlv@mail.xidian.edu.cn) for her valuable advice and help.

\ifCLASSOPTIONcaptionsoff
  \newpage
\fi



\bibliographystyle{IEEEtran}
\bibliography{bowen_bare_jrnl_physical_attack_to-anti-spoofing_CNN}
\end{document}